# Ternary Antmonides $RE_2Ir_3Sb_4$ ($RE$ = La, Ce, Pr, Nd)

Konrad Schäfer, Wilfried Hermes, Ute Ch. Rodewald, Rolf-Dieter Hoffmann, and Rainer Pöttgen

Institut für Anorganische und Analytische Chemie, Universität Münster, Corrensstrasse 30, 48149 Münster, Germany

Reprint requests to R. Pöttgen. E-mail: pottgen@uni-muenster.de

The antimonides $RE_2Ir_3Sb_4$ ($RE$ = La, Ce, Pr, Nd) were synthesized by arc-melting of the elements and subsequent annealing or *via* high-frequence melting. The samples were characterized by X-ray powder diffraction and the four structures were refined from single-crystal X-ray diffraction data: $Pr_2Ir_3Sb_4$-type, *Pnma*, $Z$ = 4, $a$ = 1621.9(2), $b$ = 458.60(8), $c$ = 1099.8(1) pm, $wR2$ = 0.036, 1558 $F^2$ values for $La_2Ir_3Sb_4$, $a$ = 1616.6(8), $b$ = 456.5(2), $c$ = 1094.8(5) pm, $wR2$ = 0.092, 1080 $F^2$ values for $Ce_2Ir_3Sb_4$, $a$ = 1613.0(5), $b$ = 454.9(2), $c$ = 1094.1(5) pm, $wR2$ = 0.057, 1428 $F^2$ values for $Pr_2Ir_3Sb_4$, and $a$ = 1609.8(6), $b$ = 452.9(2), $c$ = 1092.3(5) pm, $wR2$ = 0.052, 1472 $F^2$ values for $Nd_2Ir_3Sb_4$, with 56 parameters per refinement. The Sb1 atoms show enhanced displacement off the mirror planes at $y$ = 1/4 and $y$ = 3/4. A series of temperature dependent structure refinements of $Pr_2Ir_3Sb_4$ down to 90 K are indicative of static disorder. The iridium and antimony atoms build up complex covalently bonded three-dimensional $[Ir_3Sb_4]$ networks with Ir–Sb distances ranging from 256–269 pm ($Nd_2Ir_3Sb_4$). The two crystallographically independent rare earth sites fill cavities of coordination numbers 17 (8 Ir + 9 Sb) and 15 (6 Ir + 9 Sb) within the $[Ir_3Sb_4]$ polyanions. Temperature dependent magnetic susceptibility measurements indicate a stable trivalent ground state for the cerium compound. No magnetic ordering was evident down to 3 K.

*Key words*:   Rare Earth, Intermetallics, Crystal Structure, Magnetic Properties

**Introduction**

Antimonide phases with compositions close to Ce$T_2$Sb$_2$ ($T$ = transition metal) have repeatedly been studied in the last years with respect to their crystal structures and magnetic properties. In contrast to the many silicides and germanides with the well known highly symmetric ThCr$_2$Si$_2$- and CaBe$_2$Ge$_2$-type structures [1, 2], the antimonides show more complicated crystal chemical behavior. CeNi$_{2-x}$Sb$_2$ and CeCu$_{2-x}$Sb$_2$ with CaBe$_2$Ge$_2$-related structures show substantial defects on the transition metal sites [3-10]. The vacancy formation strongly influences the physical properties of these two antimonides. Furthermore one observes solid solutions. An interesting example is the nickel-rich phase CeNi$_{2+x}$Sb$_{2-x}$ ($x$ = 0.35) which shows an orthorhombic distortion [11]. Even more complicated is the solid solution Ce(Ag$_x$Cu$_{1-x}$)$_{2-\delta}$Sb$_2$ [12, 13], where the Néel temperature first decreases from CeCu$_{2-\delta}$Sb$_2$ ($T_N$ = 5 K) to Ce(Ag$_x$Cu$_{1-x}$)$_{2-\delta}$Sb$_2$ ($T_N$ = 3 K) and then increases towards CeAg$_{2-\delta}$Sb$_2$ ($T_N$ = 8.93 K) [13].

With the 4$d$ transition metals, besides CeAg$_{2-\delta}$Sb$_2$ [13], CePd$_2$Sb$_2$ [14] with presumably CaBe$_2$Ge$_2$-type structure and CeRh$_2$Sb$_2$ have been reported. For CeRh$_2$Sb$_2$, the assignment of the correct structure type (body-centered ThCr$_2$Si$_2$ vs. primitive CaBe$_2$Ge$_2$) is still controversely discussed [15-17]. Recent phase analytical studies in the Ce–Rh–Sb system revealed the existence of Ce$_8$Rh$_{17}$Sb$_{14}$ ≡ CeRh$_{2.13}$Sb$_{1.75}$ [18] and the isotypic antimonides $RE_8$Rh$_{17}$Sb$_{14}$ ($RE$ = La, Pr, Nd) which crystallize with a CaBe$_2$Ge$_2$-related superstructure through rhodium–antimony-ordering and vacancy formation. Similar crystal chemical behavior has been observed for the antimonide CePt$_{2+x}$Sb$_{2-y}$ [19].

In the course of our systematic studies of intermetallic 1-2-2 compounds in the field of pnictide superconductors [20-24, and refs. therein], we have been searching for related compounds in the $RE$–Ir–Sb systems. These investigations resulted in the antimon-ides $RE_2Ir_3Sb_4$ ($RE$ = La, Ce, Pr, Nd) reported herein. After completion of our work we got aware of independent work on $La_2Ir_3Sb_4$ and $Ce_2Ir_3Sb_4$ [25].

**Experimental**

*Synthesis*

Starting materials for the synthesis of the $RE_2Ir_3Sb_4$ ($RE$ = La, Ce, Pr, Nd) samples were ingots of the rare earth elements (Johnson Matthey or smart elements), iridium powder (Degussa-Hüls) and antimony lumps (Johnson Matthey), all with stated purities better than 99.9 %. In the first step, pieces of the rare earth ingots were arc-melted [26] under argon (ca. 700 mbar) to small buttons. The argon was purified before with molecular sieves, silica gel, and titanium sponge (900 K). For the synthesis of $La_2Ir_3Sb_4$, $Ce_2Ir_3Sb_4$, and $Pr_2Ir_3Sb_4$ the elements were then weighed in the atomic ratio 2 : 3 : 4 and arc-welded in tantalum ampoules. The tantalum tubes were placed in a water-cooled sample chamber of an induction furnace (Hüttinger Elektronik, Freiburg, Typ TIG 1.5/300) [27]. The ampoules were rapidly heated to 1970 K and kept at that temperature for 5 min, followed by cooling to 670 K at a rate of ca. 15 K / min. The temperature was controlled through a Sensor Therm Metis MS09 pyrometer with an accuracy of ±30 K. The samples were then quenched by switching off the power supply. The silvery brittle samples could easily be separated from the tantalum tubes. No reaction with the container material was evident.

$Nd_2Ir_3Sb_4$ was synthesized from a 2 : 3 : 4 sample (neodymium button, cold-pressed pellet of iridium, antimony pieces) by arc-melting. The product button was remelted three times to ensure homogeneity. Finally the sample was sealed in an evacuated silica tube and annealed for four weeks at 1070 K in a muffle furnace. All polycrystalline $RE_2Ir_3Sb_4$ samples are stable in air over months. Single crystals exhibit metallic luster while ground powder is dark gray.

*EDX data*

Semiquantitative EDX analyses of the crystals investigated on the diffractometer were carried out by use of a Zeiss EVO® MA10 scanning electron microscope in variable pressure mode with the rare earth trifluorides, $CeO_2$, iridium and antimony as stan-dards. The experimentally observed compositions were all close to the ideal one. No impurity elements have been observed.

*X-Ray diffraction*

The polycrystalline $RE_2Ir_3Sb_4$ samples were characterized by Guinier powder patterns (Cu$K\alpha_1$ radiation, α-quartz: $a$ = 491.30, $c$ = 540.46 pm as internal standard). The Guinier camera was equipped with an imaging plate technique (Fuji-film, BAS-READER 1800). The orthorhombic lattice parameters (Table 1) were refined using a least-squares routine. To ensure correct indexing, the experimental patterns were compared to calculated ones [28] using the refined atomic positions (*vide infra*).

Small crystalline fragments were selected from the crushed annealed $RE_2Ir_3Sb_4$ samples. The suitability for intensity data collections was first checked by Laue photographs on a Buerger camera (white Mo radiation). Data sets of good quality crystals were collected at room temperature using a STOE IPDS-II image plate system (graphite monochromatized Mo$K\alpha$ radiation; λ = 71.073 pm) in oscillation mode. For the $Pr_2Ir_3Sb_4$ crystal additional measurements were conducted down to 90 K. Numerical absorption corrections were applied to all data sets. The relevant crystallographic data and details for the data collections and evaluations are listed in Table 2.

*Structure refinements*

The four data sets showed primitive orthorhombic lattices and the systematic extinctions were compatible with space group *Pnma*. The $Pr_2Ir_3Sb_4$ structure was determined first. The starting atomic parameters were deduced from Direct Methods with SHELXS-97 [29], and the structure was refined using SHELXL-97 [30] (full-matrix least-squares on $F^2$) with anisotropic atomic displacement parameters for all sites. Due to the large difference in scattering power between iridium and antimony there was no ambiguity for the site assignments. This structural setting was then applied to the other data sets. As a check for the correct composition, all occupancy parameters were refined in a separate series of least-squares cycles. All sites were fully occupied within two standard deviations and in the final cycles the ideal occupancy parameters were assumed again. For all four crystals, the $U_{22}$ parameters of the Sb1 sites were enhanced. Since this might be indicative for a violation of the mirror planes at $y$ = 1/4 and $y$ = 3/4, we also refined the structures in the non-centrosymmetric subgroup $Pn2_1a$, however, the anisotropic displacements remained. Since the crystals showed no superstructure reflections, we additionally collected several data sets of a $Pr_2Ir_3Sb_4$ crystal down to 90 K. The refinements confirmed the original structural model, however with decreasing displacement parameters. This structural feature is discussed below. The final difference Fourier synthesis revealed no significant residual peaks (Table 2). The atomic parameters, anisotropic displacement parameters (exemplarily for the 90 and 300 K data for $Pr_2Ir_3Sb_4$) and interatomic distances (exemplarily for $Nd_2Ir_3Sb_4$) are listed in Tables 3–5.

Further details of the structure refinements may be obtained from the Fachinformationszentrum Karlsruhe, D-76344 Eggenstein-Leopoldshafen (Germany), by quoting the Registry No's. CSD–423170 ($La_2Ir_3Sb_4$), CSD–423171 ($Ce_2Ir_3Sb_4$), CSD–423172 ($Pr_2Ir_3Sb_4$, 300 K data), CSD–423173 ($Pr_2Ir_3Sb_4$, 90 K data), and CSD–423174 ($Nd_2Ir_3Sb_4$).

*Magnetic measurements*

Magnetization measurements were performed by using a VSM (Quantum Design Physical-Property-Measurement-System) in the temperature range 3–305 K and applied fields up to 1 T.

**Discussion**

*Crystal chemistry*

The four antimonides $RE_2Ir_3Sb_4$ ($RE$ = La, Ce, Pr, Nd) crystallize with their own structure type. With samarium as rare earth component already another structure with similar composition forms [31]. The cell volumes of the $RE_2Ir_3Sb_4$ antimonides decrease from the lanthanum to the neodymium compound as expected from the lanthanoid contraction. The volume of the cerium compound smoothly fits between the lanthanum and the praseodymium compound, already suggesting trivalent cerium, what has been confirmed by magnetic susceptibility measurements (*vide infra*).

In the following discussion, when we quote interatomic distances we refer to the neodymium compound. The shortest interatomic distances in $Nd_2Ir_3Sb_4$ occur for Ir–Sb. Each of the three crystallographically independent iridium atoms has between 4 and 5 antimony neighbors at Ir–Sb distances ranging from 256 to 269 pm, in excellent agree-ment with the sum of the covalent radii [32] of 267 pm. We can therefore safely assume strong covalent Ir–Sb bonding. The IrSb$_4$ and IrSb$_5$ units are condensed *via* the antimony atoms, leading to the complex three-dimensional [Ir$_3$Sb$_4$] polyanionic network emphasized in Fig. 1. Within this network one observes rows of condensed Ir$_2$Sb$_2$ rhombs which extend in the *y* direction in a ladder-like manner. The single Ir$_2$Sb$_2$ rhomb in the center of the unit cell also occurs in the structure of equiatomic CeIrSb [33, 34], where the range of Ir–Sb distances is little broader, *i. e*. 268–283 pm.

The Ir2 and Ir3 atoms have slightly distorted trigonal prismatic coordination by two antimony and four neodymium atoms (Fig. 2). These prisms are condensed *via* common neodymium edges, forming chains that extend in the *x* direction with an up-up-down-down sequence for the antimony edges. Similar chains occur in the many structurally closely related TiNiSi-type phases [35] like CeIrSb [33, 34], however, the chain sequence is a shorter up-down.

Within the [Ir$_3$Sb$_4$] polyanion the two crystallographically independent neodymium sites fill larger cavities of coordination numbers 17 (8 Ir + 9 Sb) and 15 (6 Ir + 9 Sb). The neodymium atoms are well separated from each other. The shortest distance (Nd2–Nd2) of 431 pm is much longer than the Nd–Nd distances (6 × 367 and 6 × 369 pm) in hexagonal neodymium with ABAC stacking sequence [36] and they are also well above the Hill limit for *f* electron localization [37].

Besides the strong Ir–Sb bonds, also some secondary interactions can be expected within the [Ir$_3$Sb$_4$] polyanion. The Ir1 atoms have one closest Ir1 neighbor at 299 pm, only slightly longer than in *fcc* iridium (272 pm) [36]. Also a weak contact occurs between the Sb2 and Sb4 atoms. The Sb2–Sb4 distance of 329 pm is in between the intralayer (291 pm) and interlayer (336 pm) Sb–Sb distances in elemental antimony with arsenic-type structure [36].

A peculiar bonding situation occurs for the Sb1 atoms which have almost trigonal planar iridium coordination: Ir1–Sb1–Ir3 125.7 °, Ir2–Sb1–Ir3 111.7 ° and Ir1–Sb1–Ir2 122.6 °. As is evident from Fig. 3, the Sb1 atoms have a sandwich-like coordination by two slightly distorted Pr$_2$Sb$_3$ pentagons. Thus, one observes a somewhat open coordination sphere towards the pentagons. In all four compounds the Sb1 atoms show a pro-nounced displacement parameter in this direction. The room temperature data sets showed no additional reflections and also refinements in the non-centrosymmetric group *Pn*2$_1$*a* (removal of the mirror plane) did not resolve these displacements. Also the low-temperature data sets for the Pr$_2$Ir$_3$Sb$_4$ crystal revealed no superstructure reflections. The temperature dependence of the anisotropic displacement parameters is shown in Fig. 4. Extrapolation of the $U_{11}$ and $U_{33}$ values to 0 K results in vanishing values, while a displacement component of ca. 25 pm$^2$ remains for $U_{22}$, indicative for static disorder for the discussed temperature range. Similar thermal behavior is expected also for the other crystals. A potential phase transition might take place at much lower temperature.

*Magnetic properties*

The temperature dependence of the reciprocal magnetic susceptibility of Ce$_2$Ir$_3$Sb$_4$ is presented in Figure 5. Above 150 K we observe a Curie-Weiss behavior with an experimental effective magnetic moment of 2.56(1) µ$_B$ / Ce atom, in good agreement with the free ion value of 2.54 µ$_B$ for Ce$^{3+}$. Extrapolation of the $\chi^{-1}$ vs T data to $\chi^{-1}$ = 0 led to a Weiss constant of –25.3(2) K, indicative for antiferromagnetic interactions. No magnetic ordering down to 3 K was observed.

*Acknowledgements*

This work was financially supported by the Deutsche Forschungsgemeinschaft through SPP 1458 *Hochtemperatursupraleitung in Eisenpnictiden*.

Table 1. Lattice parameters (Guinier powder data) of the ternary antimonides $RE_2Ir_3Sb_4$.

| Compound | $a$ (pm) | $b$ (pm) | $c$ (pm) | $V$ (nm$^3$) |
|---|---|---|---|---|
| $La_2Ir_3Sb_4$ | 1621.9(2) | 458.60(8) | 1099.8(1) | 0.8180 |
| $Ce_2Ir_3Sb_4$ | 1616.6(8) | 456.5(2) | 1094.8(5) | 0.8079 |
| $Pr_2Ir_3Sb_4$ | 1613.0(5) | 454.9(2) | 1094.1(5) | 0.8028 |
| $Pr_2Ir_3Sb_4$[a] | 1609.7(3) | 453.49(9) | 1091.4(2) | 0.7967 |
| $Nd_2Ir_3Sb_4$ | 1609.8(6) | 452.9(2) | 1092.3(5) | 0.7964 |

[a] Single crystal diffractometer data at 90 K.

Table 2. Crystal data and structure refinement for $RE_2Ir_3Sb_4$ (space group $Pnma$, Z = 4).

| Empirical formula | $La_2Ir_3Sb_4$ | $Ce_2Ir_3Sb_4$ | $Pr_2Ir_3Sb_4$ | $Pr_2Ir_3Sb_4$ | $Nd_2Ir_3Sb_4$ |
|---|---|---|---|---|---|
| Temperature, K | 295 | 295 | 300 | 90 | 295 |
| Unit cell dimensions | Table 1 | Table 1 | Table 1 | Table 1 | Table 1 |
| Molar mass, g mol$^{-1}$ | 1341.42 | 1343.84 | 1345.42 | 1345.42 | 1352.08 |
| Calculated density, g cm$^{-3}$ | 10.89 | 11.05 | 11.13 | 11.22 | 11.28 |
| Crystal size, $\mu$m$^3$ | 10 × 10 × 10 | 10 × 20 × 20 | 20 × 30 × 30 | 20 × 30 × 30 | 20 × 50 × 60 |
| Transm. ratio (max / min) | 0.450 / 0.073 | 0.536 / 0.243 | 0.412 / 0.138 | 0.517 / 0.178 | 0.485 / 0.103 |
| Absorption coeff., mm$^{-1}$ | 71.7 | 73.3 | 74.5 | 75.1 | 75.9 |
| Detector distance, mm | 80 | 70 | 80 | 80 | 80 |
| Exposure time, min | 5 | 8 | 8 | 5 | 8 |
| $\omega$ range / increment, deg. | 0–180 / 1.0 | 0–180 / 1.0 | 0–180 / 1.0 | 0–180 / 1.0 | 0–180 / 1.0 |
| Integr. param. A / B / EMS | 12.8 / 2.9 / 0.012 | 12.8 / 2.8 / 0.012 | 11.1 / 1.0 / 0.020 | 14.0 / 2.0 / 0.030 | 14.0 / 2.0 / 0.010 |
| $F$(000), e | 2196 | 2204 | 2212 | 2212 | 2220 |
| $\theta$ range, deg. | 2 – 32 | 2 – 28 | 2 – 31 | 2 – 31 | 2 – 32 |
| Range in $hkl$ | ±24, ±6, ±16 | ±21, ±6, ±14 | ±23, ±6, ±15 | ±23, ±6, ±15 | ±23, ±6, ±16 |
| Total no. reflections | 9906 | 5154 | 9259 | 7321 | 6484 |
| Independent reflections / $R_{int}$ | 1558 / 0.053 | 1080 / 0.188 | 1428 / 0.052 | 1419 / 0.057 | 1472 / 0.043 |
| Reflections with $I \geq 2\sigma(I)$ / $R_\sigma$ | 1273 / 0.040 | 578 / 0.186 | 908 / 0.080 | 1193 / 0.044 | 1282 / 0.031 |
| Data / parameters | 1558 / 56 | 1080 / 56 | 1428 / 56 | 1419 / 56 | 1472 / 56 |
| Goodness-of-fit on $F^2$ | 0.936 | 0.876 | 0.749 | 1.071 | 1.050 |
| R1 / wR2 for $I \geq 2\sigma(I)$ | 0.025 / 0.035 | 0.068 / 0.078 | 0.027 / 0.054 | 0.031 / 0.055 | 0.025 / 0.050 |
| R1 / wR2 for all data | 0.038 / 0.036 | 0.148 / 0.092 | 0.053 / 0.057 | 0.045 / 0.057 | 0.034 / 0.052 |
| Extinction coefficient | 0.00060(2) | 0.000004(14) | 0.00061(3) | 0.00044(4) | 0.00029(4) |
| Largest diff. peak / hole, e Å$^{-3}$ | 2.39 / –2.76 | 3.35 / –4.26 | 2.99 / –3.57 | 3.27 / –3.28 | 2.61 / –2.99 |

Table 3. Atomic coordinates and isotropic displacement parameters (pm$^2$) of La$_2$Ir$_3$Sb$_4$, Ce$_2$Ir$_3$Sb$_4$, Pr$_2$Ir$_3$Sb$_4$, and Nd$_2$Ir$_3$Sb$_4$. All atoms lie on Wyckoff sites 4$c$ ($x$, 1/4, $z$). $U_{eq}$ is defined as one third of the trace of the orthogonalized $U_{ij}$ tensor.

| Atom | $x$ | $z$ | $U_{eq}$ |
| --- | --- | --- | --- |
| **La$_2$Ir$_3$Sb$_4$ (295 K)** | | | |
| La1 | 0.00582(3) | 0.75018(6) | 89(1) |
| La2 | 0.25068(3) | 0.91285(5) | 76(1) |
| Ir1 | 0.05637(2) | 0.03441(4) | 74(1) |
| Ir2 | 0.09227(2) | 0.44291(4) | 75(1) |
| Ir3 | 0.31926(2) | 0.22260(4) | 70(1) |
| Sb1 | 0.15190(3) | 0.22468(7) | 111(1) |
| Sb2 | 0.20819(3) | 0.61516(6) | 73(1) |
| Sb3 | 0.39498(3) | 0.43809(6) | 74(1) |
| Sb4 | 0.43191(3) | 0.05034(6) | 72(1) |
| **Ce$_2$Ir$_3$Sb$_4$ (295 K)** | | | |
| Ce1 | 0.0056(2) | 0.7508(4) | 91(7) |
| Ce2 | 0.2510(2) | 0.9117(3) | 99(6) |
| Ir1 | 0.0563(1) | 0.0336(2) | 63(5) |
| Ir2 | 0.0925(1) | 0.4434(2) | 78(5) |
| Ir3 | 0.3180(2) | 0.2233(2) | 79(5) |
| Sb1 | 0.1508(3) | 0.2250(4) | 123(9) |
| Sb2 | 0.2101(2) | 0.6138(4) | 81(8) |
| Sb3 | 0.3938(3) | 0.4379(4) | 94(9) |
| Sb4 | 0.4318(2) | 0.0528(4) | 87(9) |
| **Pr$_2$Ir$_3$Sb$_4$ (300 K)** | | | |
| Pr1 | 0.00673(7) | 0.75059(11) | 77(2) |
| Pr2 | 0.25147(6) | 0.91115(9) | 61(2) |
| Ir1 | 0.05598(4) | 0.03349(8) | 50(2) |
| Ir2 | 0.09248(4) | 0.44282(7) | 54(2) |
| Ir3 | 0.31771(5) | 0.22305(8) | 50(2) |
| Sb1 | 0.15080(8) | 0.22442(14) | 102(3) |
| Sb2 | 0.21050(7) | 0.61293(13) | 51(2) |
| Sb3 | 0.39344(7) | 0.43780(13) | 52(2) |
| Sb4 | 0.43170(7) | 0.05206(13) | 51(3) |
| **Pr$_2$Ir$_3$Sb$_4$ (90 K)** | | | |
| Pr1 | 0.00680(5) | 0.75065(7) | 26(1) |
| Pr2 | 0.25157(4) | 0.91075(7) | 17(1) |
| Ir1 | 0.05587(3) | 0.03327(5) | 18(1) |
| Ir2 | 0.09241(3) | 0.44308(5) | 20(1) |
| Ir3 | 0.31741(3) | 0.22299(5) | 16(1) |
| Sb1 | 0.15041(6) | 0.22445(10) | 41(2) |
| Sb2 | 0.21068(5) | 0.61255(9) | 17(2) |
| Sb3 | 0.39339(5) | 0.43758(9) | 17(2) |
| Sb4 | 0.43162(5) | 0.05247(9) | 21(2) |
| **Nd$_2$Ir$_3$Sb$_4$ (295 K)** | | | |
| Nd1 | 0.00671(4) | 0.75111(5) | 96(1) |
| Nd2 | 0.25159(3) | 0.91112(5) | 75(1) |
| Ir1 | 0.05600(3) | 0.03378(3) | 60(1) |
| Ir2 | 0.09263(3) | 0.44311(3) | 71(1) |
| Ir3 | 0.31741(3) | 0.22342(3) | 66(1) |
| Sb1 | 0.15047(5) | 0.22490(6) | 121(2) |
| Sb2 | 0.21112(4) | 0.61266(6) | 73(1) |
| Sb3 | 0.39319(5) | 0.43774(6) | 72(1) |
| Sb4 | 0.43172(4) | 0.05295(6) | 68(1) |

Table 4. Anisotropic displacement parameters ($U_{ij}$ in pm$^2$) for Pr$_2$Ir$_3$Sb$_4$ at 300 and 90 K. The anisotropic displacement factor exponent takes the form: $-2\pi^2[(ha^*)^2U_{11} + ... + 2hka^*b^*U_{12}]$. $U_{12} = U_{23} = 0$.

| Atom | $U_{11}$ | $U_{22}$ | $U_{33}$ | $U_{13}$ |
|---|---|---|---|---|
| **300 K** | | | | |
| Pr1 | 96(5) | 71(4) | 65(5) | −4(4) |
| Pr2 | 60(4) | 62(4) | 62(5) | −13(4) |
| Ir1 | 42(3) | 54(3) | 55(4) | −3(3) |
| Ir2 | 44(3) | 60(3) | 59(4) | −2(3) |
| Ir3 | 60(3) | 51(3) | 39(4) | 2(3) |
| Sb1 | 45(5) | 210(7) | 50(7) | −3(5) |
| Sb2 | 70(5) | 36(5) | 47(6) | −7(4) |
| Sb3 | 49(5) | 43(5) | 65(7) | −13(4) |
| Sb4 | 34(5) | 53(5) | 65(7) | 9(4) |
| **90 K** | | | | |
| Pr1 | 30(3) | 21(2) | 27(3) | −11(2) |
| Pr2 | 16(3) | 14(3) | 22(3) | −4(3) |
| Ir1 | 22(2) | 14(2) | 19(2) | 0(2) |
| Ir2 | 18(2) | 17(2) | 24(3) | 3(2) |
| Ir3 | 26(2) | 10(2) | 12(3) | 0(2) |
| Sb1 | 17(4) | 86(3) | 19(4) | −1(3) |
| Sb2 | 24(4) | 3(3) | 25(4) | −8(3) |
| Sb3 | 15(4) | 10(3) | 27(4) | −4(3) |
| Sb4 | 22(4) | 14(3) | 26(4) | 6(3) |

Table 5. Interatomic distances (pm), calculated with the powder lattice parameters of $Nd_2Ir_3Sb_4$. All distances within the first coordination spheres are listed. Standard deviations are all equal or less than 0.2 pm.

| Nd1: | 1 | Ir1 | 318.8 | Ir3: | 2 | Sb2 | 260.8 |
|---|---|---|---|---|---|---|---|
|  | 2 | Sb4 | 328.6 |  | 1 | Sb4 | 261.8 |
|  | 2 | Sb1 | 340.6 |  | 1 | Sb3 | 264.0 |
|  | 2 | Ir1 | 341.6 |  | 1 | Sb1 | 268.7 |
|  | 2 | Sb3 | 344.7 |  | 2 | Nd2 | 325.0 |
|  | 2 | Ir2 | 349.1 |  | 1 | Nd2 | 357.2 |
|  | 1 | Sb4 | 353.4 |  | 2 | Nd1 | 363.8 |
|  | 1 | Sb2 | 362.1 | Sb1: | 1 | Ir2 | 255.9 |
|  | 1 | Ir2 | 363.8 |  | 1 | Ir1 | 258.3 |
|  | 2 | Ir3 | 363.8 |  | 1 | Ir3 | 268.7 |
|  | 1 | Sb3 | 385.9 |  | 2 | Sb2 | 340.5 |
| Nd2: | 2 | Sb2 | 321.5 |  | 2 | Nd1 | 340.6 |
|  | 2 | Ir3 | 325.0 |  | 2 | Nd2 | 342.8 |
|  | 2 | Sb3 | 326.3 |  | 1 | Nd2 | 379.4 |
|  | 1 | Sb4 | 328.8 | Sb2: | 2 | Ir3 | 260.8 |
|  | 1 | Sb2 | 332.5 |  | 1 | Ir2 | 265.9 |
|  | 2 | Ir2 | 339.7 |  | 2 | Nd2 | 321.5 |
|  | 1 | Ir1 | 342.2 |  | 2 | Sb4 | 329.3 |
|  | 2 | Sb1 | 342.8 |  | 1 | Nd2 | 332.5 |
|  | 1 | Ir3 | 357.2 |  | 2 | Sb1 | 340.5 |
|  | 1 | Sb1 | 379.4 |  | 1 | Sb3 | 349.9 |
| Ir1: | 1 | Sb1 | 258.3 |  | 1 | Nd1 | 362.1 |
|  | 2 | Sb3 | 262.6 | Sb3: | 2 | Ir1 | 262.6 |
|  | 1 | Sb3 | 263.9 |  | 1 | Ir1 | 263.9 |
|  | 2 | Ir1 | 298.7 |  | 1 | Ir3 | 264.0 |
|  | 1 | Nd1 | 318.8 |  | 2 | Nd2 | 326.3 |
|  | 2 | Nd1 | 341.6 |  | 2 | Nd1 | 344.7 |
|  | 1 | Nd2 | 342.2 |  | 1 | Sb2 | 349.9 |
| Ir2: | 1 | Sb1 | 255.9 |  | 1 | Nd1 | 385.9 |
|  | 1 | Sb4 | 259.1 | Sb4: | 1 | Ir2 | 259.1 |
|  | 2 | Sb4 | 259.3 |  | 2 | Ir2 | 259.3 |
|  | 1 | Sb2 | 265.9 |  | 1 | Ir3 | 261.8 |
|  | 2 | Nd2 | 339.7 |  | 2 | Nd1 | 328.6 |
|  | 2 | Nd1 | 349.1 |  | 1 | Nd2 | 328.8 |
|  | 1 | Nd1 | 363.8 |  | 2 | Sb2 | 329.3 |
|  |  |  |  |  | 2 | Sb4 | 336.1 |
|  |  |  |  |  | 1 | Nd1 | 353.4 |

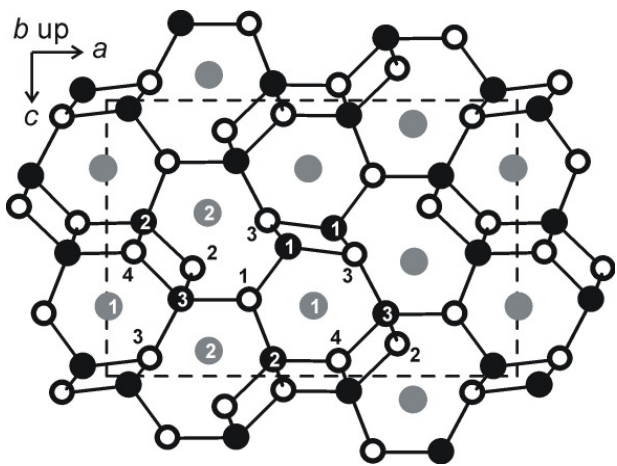

Fig. 1. Projection of the $Nd_2Ir_3Sb_4$ structure along the short unit cell axis. Neodymium, iridium, and antimony atoms are drawn as medium gray, black filled and open circles, respectively. The three-dimensional [$Ir_3Sb_4$] network is emphasized.

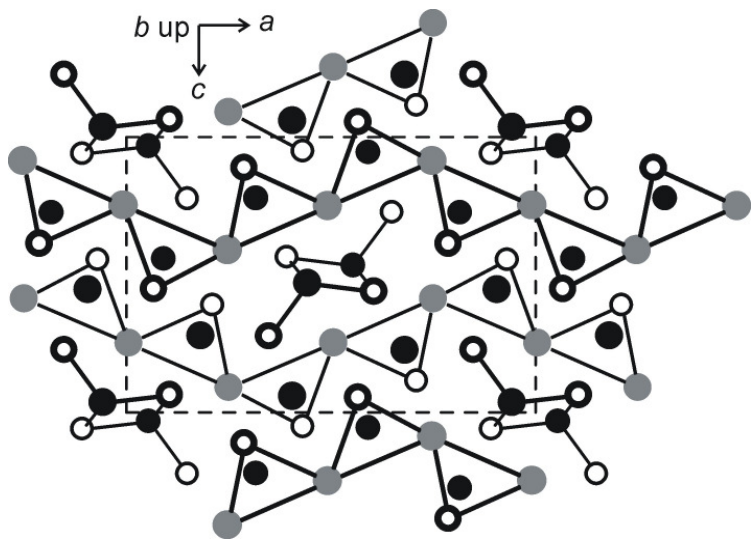

Fig. 2. Projection of the $Nd_2Ir_3Sb_4$ structure along the short unit cell axis. Neodymium, iridium, and antimony atoms are drawn as medium gray, black filled and open circles, respectively. All atoms lie on mirror planes at $y = 1/4$ (thin lines) and $y = 3/4$ (thick lines. The trigonal prismatic iridium coordination is emphasized.

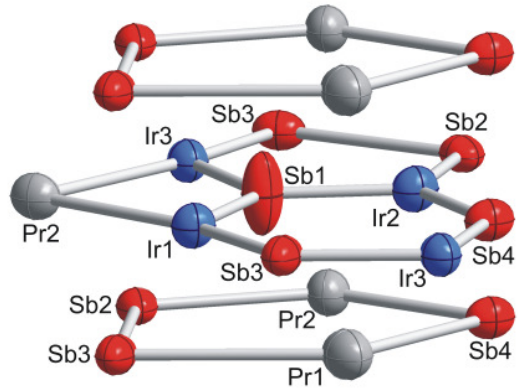

Fig. 3 (color online). Coordination of the Sb1 atoms in $Pr_2Ir_3Sb_4$. Displacement ellipsoids are drawn at the 99 % probability level.

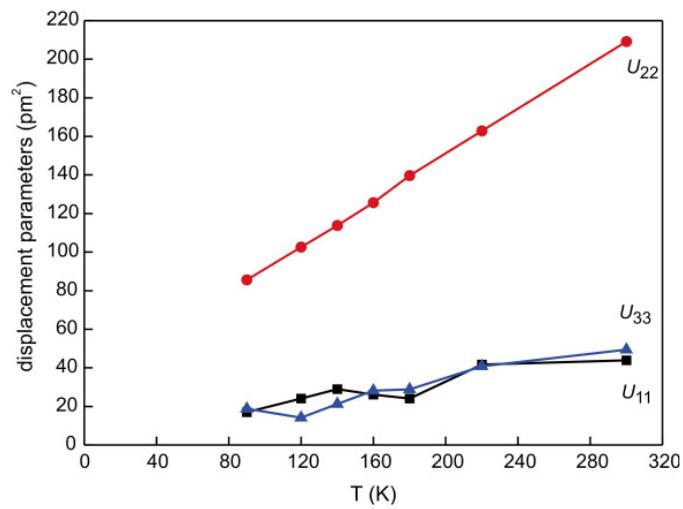

Fig. 4 (color online). Temperature dependence of the anisotropic displacement parameters $U_{11}$, $U_{22}$, and $U_{33}$ of $Pr_2Ir_3Sb_4$. For details see text.

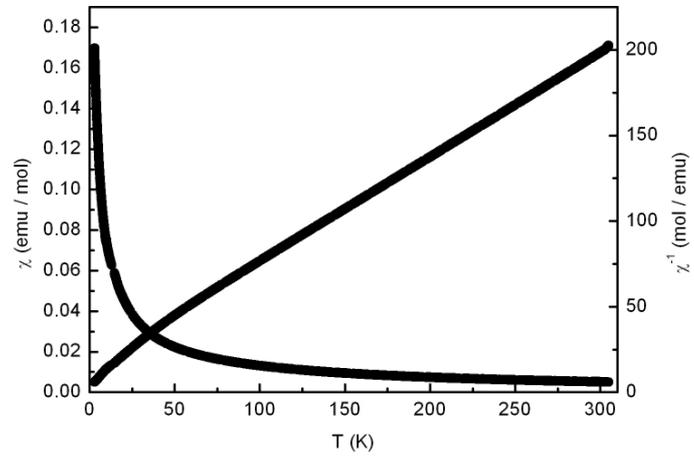

Fig. 5. Temperature dependence of the magnetic susceptibility ($\chi$ and $\chi^{-1}$ data) of $Ce_2Ir_3Sb_4$ measured at 10 kOe.